\documentclass{JAC2003}
\usepackage{graphicx}

\begin{document}
\title{PHYSICS WITH CHARGED KAONS: \\ RECENT AND FUTURE EXPERIMENTS}

\author{M. S. Sozzi, CERN, Geneva, Switzerland\thanks{On leave from Scuola
Normale Superiore, Pisa. E-mail: \texttt{marco.sozzi@cern.ch}}}

\maketitle

\vspace{0.5cm}

\begin{abstract}
This paper summarizes some recent progress and future perspectives in the 
experimental investigation of the Standard Model (SM) (and physics beyond it) 
using charged kaon decays, except for the important mode 
$K^\pm \rightarrow \pi^\pm \nu \overline{\nu}$ which is discussed in detail 
elsewhere \cite{pinunu}.
\end{abstract}

\section{INTRODUCTION}
Kaons, being the minimal ``flavour laboratory'', always had a leading 
r\^ole in the discovery and the study of fundamental physics issues which are 
related to flavour changing transitions induced by weak interactions. 
Their success as a physics probe is partly due to the mixed blessing of their 
relatively small mass (on the hadronic scale), and its experimental 
consequences.

Along the history from the discovery of CP violation \cite{Cronin} to the 
evidence of direct CP violation \cite{DirectCP} \cite{Cimento} the peculiar 
system of neutral kaons was central, while charged kaons were deeply exploited 
for the search of very rare processes probing high energy scales \cite{rare}.

The level of sensitivity at which rare kaon decay searches have been pushed is 
rather remarkable: several branching ratios at the $10^{-8}$ level have been
measured (not to mention the smallest branching ratio ever measured, 
$BR(K_L \rightarrow e^+ e^-) \simeq 9 \cdot 10^{-12}$), and limits down to 
$10^{-11}$ are available \cite{PDG}.

Today, great efforts are dedicated to testing the Standard Model by 
over-constraining the CKM mixing matrix, and it should be noted that also in 
this respect kaon decays might offer a very powerful probe, complementary and 
in some cases even superior to the one represented by $B$ mesons 
\cite{CKM_KB}, for the decays which can be reliably described by the theory.
We do not know where new physics will show up first, and even relatively 
well measured parameters such as the Cabibbo angle might reserve some surprise 
\cite{Vus}, so that more refined measurements of several kaon decays are now 
of great interest.

For what concerns CP violation searches, it should be noted that so far the 
only evidence of this phenomenon in Nature arises from somewhat subtle effects 
involving coupled neutral states. While after the discovery of direct CP 
violation we have no reason to doubt that CP asymmetries should be present 
also in charged particle decays, as predicted by our current paradigm for
describing CP violation, such asymmetries - which would be indeed the simplest 
manifestations of CP violation from a conceptual point of view and entirely 
due to direct CP violation - have yet to be detected (although admittedly we 
might be close to this). 
T-odd correlations, being actively probed in charged K decays through 
polarization measurements, are also a complementary window on new physics 
through CP violation (also T violation has only been detected in the state 
mixing of neutral kaons so far).

One should also mention that precise kaon decay data is needed to further 
constrain the parameters of the effective theory we use to describe 
low-energy physics, namely chiral perturbation theory, and that measurements 
of branching ratios and decay form factors can offer stringent tests of 
anomalous couplings and universality.

\section{LEPTON FLAVOUR NUMBER VIOLATION}

Several experiments were devoted to searches for lepton flavour number 
violation (LFV) by searching for rare decays of charged kaons. 
These measurements allowed to dismiss several models of new physics thanks to 
their sensitivity to very high mass scales, and represent nowadays an 
important constraint to models of physics beyond the SM.
The results achieved by several dedicated experiments, lately mainly by 
BNL experiment E865, are rather impressive (see table \ref{tab:lfv}). 
The low limits reached make further progress difficult: kaon flux and 
physical backgrounds are the limiting factors, and no new dedicated 
experiments on LFV with kaons are in preparation at this time.

\begin{table}[hbt]
\begin{center}
\caption{Present limits on LFV charged kaon decays \cite{PDG}.}
\vspace{0.3cm}
\begin{tabular}{|l|c|}
\hline
\textbf{Mode} & \textbf{BR limit (90\% CL)} \\ 
\hline
$K^+ \rightarrow \pi^+ \mu^+ e^-$   & $2.8 \cdot 10^{-11}$ \\
$K^+ \rightarrow \pi^+ \mu^- e^+$   & $5.2 \cdot 10^{-10}$ \\
$K^+ \rightarrow \pi^- \mu^+ e^+$   & $5.0 \cdot 10^{-10}$ \\
$K^+ \rightarrow \pi^- e^+ e^+$     & $6.4 \cdot 10^{-10}$ \\
$K^+ \rightarrow \pi^- \mu^+ \mu^+$ & $3.0 \cdot 10^{-9}$ \\
\hline
\end{tabular}
\label{tab:lfv}
\end{center}
\end{table}

\section{HADRONIC DECAYS AND \\ CP VIOLATION}

The decay amplitudes for $K^\pm \rightarrow 3\pi$ have been recomputed 
recently at next-to-leading order in the chiral perturbation expansion 
\cite{Bijnens_3pi} \cite{Prades_3pi}, partially accounting for 
isospin-breaking effects. 
The corrections with respect to the leading order results are found to be of 
the order 30\%, and the fitted values of the Dalitz plot slopes (see 
\cite{PDG} for the definitions) agree rather well with the experimental data, 
as shown in table \ref{tab:K3pi}.

\begin{table*}[hbt]
\begin{center}
\caption{Dalitz plot slopes for $K^\pm \rightarrow 3\pi$ decays, from 
\cite{PDG}, \cite{KLOE_3pin} and \cite{ISTRA_3pin} (the errors are inflated 
according to the PDG recipe to account for the poor consistency).}
\vspace{0.3cm}
\begin{tabular}{|l|c|c|c|}
\hline
\textbf{Parameter} & \textbf{Experiment} & $\mathbf{\chi^2/}$\textbf{dof} & 
\textbf{Theory fit} \\
\hline
$g(\pi^\pm \pi^+ \pi^-)$ &
$-0.2160 \pm 0.0029$ & 2.6 & $-0.22 \pm 0.02$ \\
$h(\pi^\pm \pi^+ \pi^-)$ & 
$0.011 \pm 0.004$    & 1.3 & $0.012 \pm 0.005$ \\
$k(\pi^\pm \pi^+ \pi^-)$ & 
$-0.0093 \pm 0.0022$ & 3.2 & $0.0054 \pm 0.0015$ \\
\hline
$g(\pi^\pm \pi^0 \pi^0)$ &
$0.628 \pm 0.019$    & 6.8 & $0.61 \pm 0.05$ \\
$h(\pi^\pm \pi^0 \pi^0)$ & 
$0.045 \pm 0.011$    & 1.8 & $0.069 \pm 0.018$ \\
$k(\pi^\pm \pi^0 \pi^0)$ & 
$0.003 \pm 0.006$    & 5.6 & $0.004 \pm 0.002$ \\
\hline 
\hline
\end{tabular}
\label{tab:K3pi}
\end{center}
\end{table*}

New results were obtained recently for the $\pi^\pm \pi^0 \pi^0$ decay 
mode: a preliminary branching ratio measurement at 1\% with small background 
was obtained by KLOE at DA$\Phi$NE \cite{KLOE_3pin}:
\begin{displaymath}
  BR(K^\pm \rightarrow \pi^\pm \pi^0 \pi^0) = (1.781 \pm 0.013 \pm 0.016)\%
\end{displaymath}
from 440 pb$^{-1}$ of integrated luminosity. The constrained kinematics with 
high-purity pion and muon tagging (obtained from $\pi^\pm \pi^0$ and 
$\mu^\pm \nu$ decays respectively) are strong points for this experiment, 
which has good prospects for improving our knowledge of the charged kaon decay 
parameters. The $8 \cdot 10^{31}$ cm$^{-2}$ s$^{-1}$ peak luminosity obtained 
in 2002 (leading to a total of 500 pb$^{-1}$ integrated luminosity) is less 
than an order of magnitude below the design one, which is expected to be 
reached with steady improvements on the machine.

New results on the Dalitz plot slopes \cite{KLOE_3pin} \cite{ISTRA_3pin}  
helped improving the experimental determination of some parameters, 
although the consistency of the data remains rather poor, as seen from 
table \ref{tab:K3pi}, particularly for what concerns the linear slope of 
$\pi^\pm \pi^0 \pi^0$ (see also \cite{KEK_3pin}).

The 30-year long quest for direct CP violation \cite{Cimento} recently 
resulted in the definitive experimental evidence of this phenomenon 
\cite{DirectCP}, ruling out the super-weak \emph{ansatz}. 
The $\epsilon'/\epsilon$ parameter describing direct CP violation in neutral 
kaon decays, however, is still under poor theoretical control (to the extent 
that the experimental value could still be saturated by new physics), so that 
more measurement of CP violation in different systems are required to really 
test the present paradigm. 
Long ago it was suggested that charge asymmetries in $K^\pm$ hadronic 
decays, such as that for the linear Dalitz plot slope parameter $g$ in the 
decay to 3 charged pions, related to the kinetic energy distribution of the 
odd pion 
\begin{displaymath}
  A_g(\pi^\pm \pi^+ \pi^-) = 
\frac{g(\pi^+ \pi^+ \pi^-) - g(\pi^- \pi^- \pi^+)}
     {g(\pi^+ \pi^+ \pi^-) + g(\pi^- \pi^- \pi^+)}
\end{displaymath}
being in principle free from the amplitude suppression due to the 
$\Delta I = 1/2$ rule, could be a valid alternative to $\epsilon'/\epsilon$ as 
a measure of direct CP violation.

From a theoretical point of view, several predictions for the asymmetries in 
charged kaon decay parameters are available in the literature 
(see \cite{Prades_3pi} \cite{Isidori_review} and references therein), spanning 
a rather wide range of values. 
At this time, however, the common understanding is that in the SM such 
asymmetries, although not fully under theoretical control, are rather tiny for 
accidental reasons generally below $10^{-4}$ (\emph{i.e.} 
$A_g(\pi^\pm \pi^+ \pi^-) \sim 0.5 \cdot 10^{-4}$ at most \cite{Prades_3pi});
searches focus on common decay modes.
In some extensions of the Standard Model somewhat larger values, above 
$10^{-4}$ could be reached \cite{Shabalin} \cite{Martinelli}.

For $3\pi$ decays the partial decay width asymmetries are suppressed with 
respect to the ones for the Dalitz plot slopes. Asymmetries are not expected 
to be significantly larger for other decay modes, such as $K^\pm \rightarrow 
\pi^\pm \ell^+ \ell^-$ ($\ell=e,\mu$), the (relatively) larger one 
($O(10^{-4}$) being probably expected to be the photon spectrum asymmetry in 
$K^\pm \rightarrow \pi^\pm \pi^0 \gamma$ decays, which were recently measured 
with high statistics \cite{KEK_pi+pi0gamma}.

The experimental information on differences in decay parameters for $K^+$ 
and $K^-$, which would be a sign of direct CP violation, is rather poor at 
present, as shown in table \ref{tab:asym}.

\begin{table}[hbt]
\begin{center}
\caption{Experimental data on CP-violating asymmetries for $K^\pm$ decay rates 
($A_\Gamma$) and Dalitz plot slope parameters ($A_g, A_h, A_k$), from 
\cite{PDG}, \cite{HyperCP_3pi}, \cite{HyperCP_pimumu}. 
In the third column ``meas'' indicates measurements, and ``PDG'' naive 
asymmetries evaluated from PDG values (using inflated errors) when no 
measurements are available.}
\vspace{0.3cm}
\begin{tabular}{|l|c|c|}
\hline
\textbf{Mode} & \textbf{Asymmetry} & \textbf{Notes} \\
\hline
$A_\Gamma(\pi^\pm \pi^+ \pi^-)$ &
$(0.07 \pm 0.12)\%$ & meas \\
$A_g(\pi^\pm \pi^+ \pi^-)$ &
$(-0.11 \pm 0.34)\%$ & meas \\
$A_h(\pi^\pm \pi^+ \pi^-)$ &
  $(9 \pm 44)\%$ & PDG \\
$A_k(\pi^\pm \pi^+ \pi^-)$ &
  $(9 \pm 20)\%$ & PDG \\
\hline
$A_\Gamma(\pi^\pm \pi^0 \pi^0)$ &
$(0.0 \pm 0.6)\%$ & meas \\
$A_g(\pi^\pm \pi^0 \pi^0)$ &
  $(5.2 \pm 2.8)\%$ & PDG \\
$A_h(\pi^\pm \pi^0 \pi^0)$ &
  $(20 \pm 21\%)$ & PDG \\
$A_k(\pi^\pm \pi^0 \pi^0)$ &
  $(90 \pm 16\%)$ & PDG \\
\hline
$A_\Gamma(\mu^\pm \nu)$ &
$(-0.54 \pm 0.41)\%$ & meas \\
$A_\Gamma(\pi^\pm \pi^0)$ &
$(0.8 \pm 1.2)\%$ & meas \\
$A_\Gamma(\pi^\pm \pi^0 \gamma)$ &
$(0.9 \pm 3.3)\%$ & meas \\
$A_\Gamma(\pi^\pm \mu^+ \mu^-)$ &
$(-2 \pm 12)\%$ & meas \\
\hline 
\hline
\end{tabular}
\label{tab:asym}
\end{center}
\end{table}


Dalitz plot slope asymmetries have been directly measured only for the 
$\pi^\pm \pi^+ \pi^-$ mode. A preliminary result was recently obtained, as a 
byproduct, by the HyperCP experiment at FNAL, devoted to the search for CP 
violation in hyperon decays.
With the 1997 data sample (about $4.2 \cdot 10^7$ $K^+$ and $1.2 \cdot 10^7$ 
$K^-$ decays, corresponding to 20\% of the total collected statistics) the 
preliminary result \cite{HyperCP_3pi}
\begin{displaymath}
  A_g(\pi^\pm \pi^+ \pi^-) = (-0.22 \pm 0.15 \pm 0.37)\%
\end{displaymath}
was obtained (the first error being statistical and the second systematic).
The largest systematics (expected to be improved in a full analysis) are given 
by the effect of residual uncontrolled magnetic fields and Monte Carlo 
corrections.

Other large differences between the measured parameters for $K^+$ and $K^-$ in 
different experiments hint at problems in the data; removing a single out-lier 
measurement all the slope asymmetries become consistent with zero. Clearly, 
new experiments measuring both kaon charge partners, to better keep 
systematics under control, are required to improve the situation.


The small SM predictions for direct CP violating asymmetries in charged
kaon decays allow for a large window of opportunity in searching for new
physics.
Two major experimental programs are being carried on to search for direct CP 
violation by measuring a charge asymmetry in the linear Dalitz plot slopes 
$g$ of $K^\pm \rightarrow 3\pi$ decays.

\begin{figure*}[htb]
\centering
\includegraphics*[width=145mm]{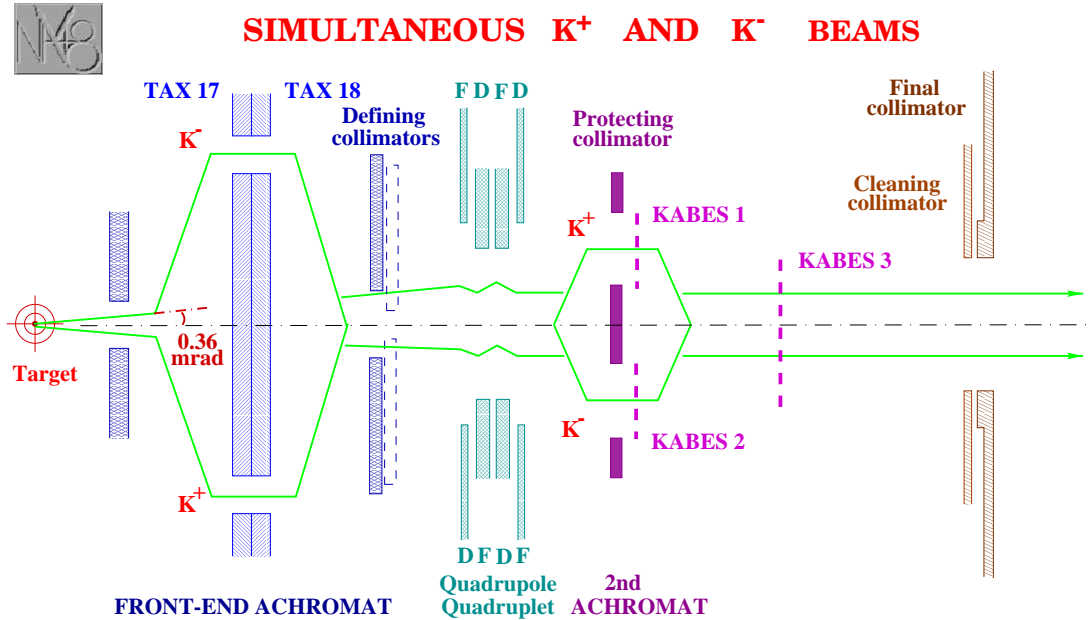}
\caption{Layout of the NA48/2 simultaneous $K^\pm$ beam line.}
\label{fig:NA48/2}
\end{figure*}

The NA48/2 experiment at the CERN SPS \cite{NA48/2} uses a new beam line 
delivering simultaneous positive and negative unseparated hadron beams of 60 
GeV/$c$ average momentum (with a $\pm$ 5\% spread) to the improved NA48 
detector.
The beams (see fig. \ref{fig:NA48/2}), produced at zero degrees by 
$7 \cdot 10^{11}$ protons per 4.8 s pulse from the SPS every 16.8 s, contain 
about 5\% $K^\pm$, resulting in $2.2(1.3) \cdot 10^6$ $K^+(K^-)$ entering the 
decay volume per pulse.

A high-rate beam spectrometer based on MICROMEGA chambers (``KABES'') allows 
the measurement of the incoming particle charge, momentum and direction, 
complementing the magnetic spectrometer downstream; most of the flux from beam 
pion decays remains in the beam pipe crossing all detectors.
By frequently switching the polarity of the analyzing and beam-line magnets, 
the experiment can cancel most of the systematics linked to asymmetries of 
the apparatus.
NA48/2 took data for the first time in 2003, and a second run in 2004 will 
largely increase the statistics, to allow reaching the design error of 
$1.7 \cdot 10^{-4}$ on $A_g(\pi^\pm \pi^+ \pi^-)$.
The experiment also collects a large amount of $K^\pm \rightarrow \pi^\pm 
\pi^0 \pi^0$ decays, to reach a similar level of accuracy on the corresponding 
slope asymmetry.

The OKA experiment \cite{OKA} is in preparation at Protvino and will be 
installed on a new RF separated kaon beam line of 15 GeV/$c$ ($\pi$
contamination $<$50\%) at the U-70 PS. Only one charge at the time will be 
available, but the high statistics ($3 \cdot 10^{13}$ protons per pulse 
giving a flux of $4(1.3) \cdot 10^6$ $K^+(K^-)$ entering the decay volume) is 
expected to allow reaching an error of $1 \cdot 10^{-4}$ on 
$A_g(\pi^\pm \pi^+ \pi^-)$.


The magnetic detector which will be used is an evolution of the ISTRA+ and 
GAMS ones.
Several rare decays will also be studied with this setup.
Half of the beam line is now ready and the first run is expected in November 
2004.

The KLOE experiment, with its source of correlated $K^+ K^-$ pairs, is also 
expected to contribute to the measurement of charge asymmetries when 
enough statistics will be available, and its results will be affected by 
rather different systematics.

\section{SEMI-LEPTONIC DECAYS}
\begin{figure}[htb]
\centering
\includegraphics*[width=65mm]{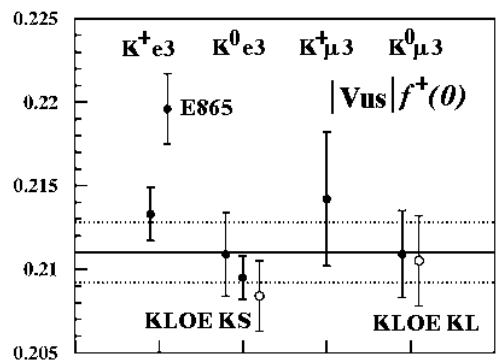}
\caption{Recent determinations of the $|V_{us}|$ element of the CKM mixing 
matrix from semi-leptonic kaon decays.} 
\label{fig:Vus}
\end{figure}

The interest in the measurement of semi-leptonic decays of kaons was revived
recently due to the present unsatisfactory agreement of experimental data with 
one of the unitarity constraints on the CKM mixing matrix.
The constraint relation $|V_{ud}|^2 + |V_{us}|^2 + |V_{td}|^2 = 1$ is 
violated at the 2.2 standard deviation level using the 2002 PDG data 
\cite{PDG}. Half of the error is due to the knowledge of the $|V_{us}|$ 
matrix element, which is extracted from hyperon and kaon semi-leptonic decays.
A new measurement of the $K^+ \rightarrow \pi^0 e^+ \nu$ ($K^+_{e3}$) 
branching ratio, obtained in a 1 week special run of experiment E865 at 
Brookhaven \cite{E865_ke3}, gives a value for $|V_{us}|$ which disagrees with 
the previous measurements, but is in agreement with the unitarity constraint. 
Preliminary results from KLOE (with 78 pb$^{-1}$) \cite{KLOE_kl3} instead 
confirm previous results and the unitarity problem (see fig. \ref{fig:Vus}).
 
New results are expected by the KLOE, NA48/2 and KTeV experiments, which 
should clarify the situation.

Another issue in semi-leptonic kaon decays concerns the experimental hints of 
the presence of scalar and/or tensor anomalous couplings in the form factors
of $K^+_{e3}$ decays, which arose several years ago.
A recent high statistics ($5.5 \cdot 10^5$ $K^-_{e3}$ decays) measurement by 
the ISTRA+ experiment at Protvino \cite{ISTRA_ke3} gave: 
\begin{eqnarray*}
  && f_S/f_+(0) = 0.002^{+0.020}_{-0.022} \pm 0.003 \\
  && f_T/f_+(0) = 0.021^{+0.064}_{-0.075} \pm 0.026 
\end{eqnarray*}
further supporting the conclusion of the E246 experiment at KEK \cite{KEK_ke3} 
which denied the existence of such anomalous couplings.

The E246 experiment also performed an interesting $\mu/e$ universality test
by comparing the $\lambda_0$ form factor slope measurement with the value 
obtained by the $K^+_{\mu3}/K^+_{e3}$ branching fraction ratio.

\section{T-ODD CORRELATIONS}
A long standing field of investigation is the search for T-odd correlations,  
such as the transverse muon polarization $P_T(\mu)$ in $K_{\mu3}$ decays.
While for the $K_L$ decay the limit due to SM final state interaction effects 
was reached in the 70's, the investigation with $K^+$ decays is still away 
from such a limit, and the tiny ($< 10^{-5}$) SM effects allow these searches 
to be a good probe for new physics.
The E246 experiment at KEK exploits the semi-muonic decays of stopped $K^+$
and forms a double ratio of events with opposite decay plane orientation to 
reduce the systematic sensitivity to spurious asymmetries.
They recently published a new (null) result \cite{E246_new} based on the 
full collected statistics ($8.3 \cdot 10^6$ $K^+_{\mu3}$ stopped decays):
\begin{displaymath}
  P_T(\mu) = (-1.12 \pm 2.17 \pm 0.90) \cdot 10^{-3} 
\end{displaymath}
and for the first time also a result for the $\mu^+ \nu \gamma$ decay
\cite{E246_munug}, which are complementary to the former mode for 
discriminating physics beyond the SM:
\begin{displaymath}
  P_T(\mu) = (-0.64 \pm 1.85 \pm 0.10) \cdot 10^{-2} 
\end{displaymath}
The experiment is now completed: its final sensitivity (for the $K^+_{\mu3}$ 
mode) is expected to be around $1.5 \cdot 10^{-3}$ in the transverse 
polarization; the main systematics arise from residual detector misalignments, 
asymmetric spurious magnetic fields and the large in-plane muon polarization.

Improved experiments on transverse muon polarization have been proposed, which 
could reach the $\sim 1 \cdot 10^{-4}$ sensitivity.

Another class of T-odd correlations is that involving only three-momenta 
in four-body final states, such as radiative semi-leptonic decays. As in the 
previous case, the small SM contributions from final state interactions leaves 
a large window of opportunity for searches of new physics. In particular, new 
sources of CP violation with vector or axial couplings are not constrained by 
the transverse muon polarization results \cite{Chalov}, and will be searched 
for in the NA48/2 and OKA experiments.

\section{RARE DECAYS}
Apart from the very important $K \rightarrow \pi \nu \overline{\nu}$ 
decays which are not discussed here, the class of FCNC loop-induced decays 
comprises also $K^\pm \rightarrow \pi^\pm \ell^+ \ell^-$ (where $\ell$ is a 
charged lepton). These are less interesting than the previous ones (or even 
their $K_L$ counterparts) since long-distance physics dominates. Still, the 
value of the ratio of $K^+ \rightarrow \pi^+ \mu^+ \mu^-$ to $K^+ \rightarrow 
\pi^+ e^+ e^-$ decays is constrained in a model-independent way in the 
chiral perturbation expansion, and the experimental situation was unclear 
recently, with two incompatible experimental results (at 3.4 standard 
deviations) for the $K^+ \rightarrow \pi^+ \mu^+ \mu^-$ branching ratio, one 
of which would violate the constraint mentioned above. 
A new preliminary measurement by the HyperCP experiment \cite{HyperCP_pimumu} 
\begin{displaymath}
  BR(K^\pm \rightarrow \pi^\pm \mu^+ \mu^-) = 
  (9.8 \pm 1.0 \pm 0.5) \cdot 10^{-8}
\end{displaymath}
seems to clarify the situation, bringing the experimental value of the above 
ratio within the limits allowed by the SM.

CP-violating asymmetries are expected to be tiny also in these modes, which 
makes them unattractive for such searches.

Among other interesting rare decay modes of charged kaons, $\pi^\pm \gamma 
\gamma$ should be mentioned, with $\sim 10^{-6}$ branching ratio. This 
channel is very interesting as a constraint on chiral perturbation theory, 
since in involves a free $O(p^4)$ parameter and important $O(p^6)$ 
contributions, complementing the information which can be extracted from the 
$K_{L,S} \rightarrow \pi^0 \gamma \gamma$ decays. Due to the large ($\times 
2 \cdot 10^5$) irreducible background from $\pi^\pm \pi^0$, this mode is 
rather penalized from the experimental point of view, but high flux 
experiments are expected to improve its knowledge.

Similarly, the 4-lepton modes $\ell^\pm \nu \ell^+ \ell^-$, two of which were 
recently observed with much improved statistics and 10-20\% background at BNL 
\cite{E865_4leptons}:
\begin{eqnarray*}
  && \hspace{-0.5cm} 
     BR(K^+ \rightarrow e^+ \nu e^+ e^-) = (2.48 \pm 0.14 \pm 0.14) 
     \cdot 10^{-8} \\
  && \hspace{-0.5cm} 
     BR(K^+ \rightarrow \mu^+ \nu e^+ e^-) = (7.06 \pm 0.16 \pm 0.26) 
     \cdot 10^{-8}
\end{eqnarray*}
(with $m_{ee}>$ 150 and 145 MeV/$c^2$ respectively), are also good probes of 
chiral perturbation predictions \cite{Bijnens_4leptons}. 
The modes with muon pairs have not been detected yet, but their branching 
ratios are predicted to be not far below the ones for the electron modes, so 
that they should be measurable in high flux experiments.

Some more exotic searches have been performed or are in the programs of future 
charged kaon experiments, such as those for supersymmetric particles or 
$K^+ \rightarrow \pi^+ \gamma$ decays (!). 

\section{$\pi \pi$ INTERACTIONS}
It is well known that $K_{e4}$ decays are one of the best places to study 
low-energy $\pi\pi$ interactions, thanks to the Fermi-Watson theorem; indeed, 
the asymmetry in the distribution of the angle between the di-pion and the 
di-lepton planes is sensitive to the $\pi\pi$ scattering phase shifts.
The value of the S-wave isoscalar scattering length parameter is a very 
precise prediction of chiral perturbation theory \cite{Colangelo}: 
$a_0^0 = 0.220 \pm 0.005$. 
Low-energy $\pi\pi$ scattering in the S-wave is particularly sensitive 
to the size of the $\langle q \overline{q} \rangle$ condensate breaking the 
chiral simmetry of QCD.
The size of this condensate is a free parameter in chiral perturbation theory, 
and one of the assumptions (giving predictive power to the theory) is that 
it is large enough to make the lowest order term in the quark masses dominate 
the mass of light mesons.

Recently, the BNL E865 experiment improved the experimental knowledge of the 
scattering length parameter with a statistics of $4 \cdot 10^5$ $K^+_{e4}$ 
decays \cite{E865_ke4}:
\begin{displaymath}
  a_0^0 = 0.216 \pm 0.013 \pm 0.003
\end{displaymath}
While this value is in good agreement with the theoretical one, its 
statistically-dominated precision is still quite larger than the theoretical 
error, and such an unusual situation in hadronic physics calls for improved 
experimental investigations. 

The NA48/2 experiment at CERN plans to collect more than $1 \cdot 10^6$ 
$K^+_{e4}$ decays in order to be able to reach an error of $\sim 0.01$ on 
the $a_0^0$ parameter.

The DIRAC experiment at CERN \cite{DIRAC} investigates the formation of the 
electro-magnetically bound state of $\pi^+ \pi^-$ (``pionium''), and from the 
study of its lifetime expects to be able to measure $|a_0^0 - a_0^2|$ to the 
5\% level.

\begin{figure}[htb]
\centering
\includegraphics*[width=65mm]{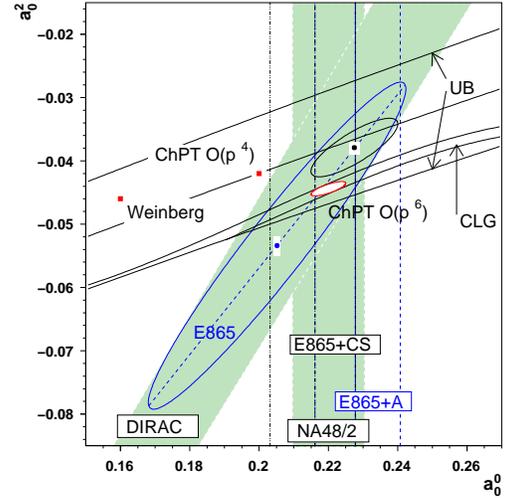}
\caption{Recent experimental determinations of the $\pi\pi$ S-wave scattering 
lengths, compared with theoretical predictions under different assumptions 
(see \cite{E865_ke4} for details); the shaded bands are estimates of the 
expected errors from DIRAC and NA48/2, for illustrative purposes.}
\label{fig:pipi}
\end{figure}

Figure \ref{fig:pipi} summarizes the current knowledge of the scattering 
length parameters.

\section{THE FUTURE}
The present scenario indicates that the future of kaon physics will most 
likely be dominated by experimental efforts devoted to overconstrain the CKM 
unitarity triangle, by measuring the very rare decay modes which can be 
predicted in the cleanest way by the theory, \emph{i.e.} $K \rightarrow 
\pi \nu \overline{\nu}$ modes \cite{pinunu}.

\begin{table*}[hbt]
\begin{center}
\caption{Present and future charged kaon sources in the world.}
\vspace{0.3cm}
\begin{tabular}{|l|l|l|l|}
\hline
\textbf{Site} & \textbf{Energy and statistics} & 
\textbf{Beam} & \textbf{Date} \\
\hline
BNL-AGS & 
6 GeV/$c$ or at rest, $> 10^{12}$ $K^+$ &
Unseparated ($\pi/K \approx 20$) or &
1995+ \\
& &
$E \times B$ separated ($\pi/K < 0.25$) & \\
KEK-PS &
At rest, $> 10^8$ $K^+$ &
$E \times B$ separated ($\pi/K \approx 6$) &
1996+ \\
LNF-DA$\Phi$NE &
100 MeV/$c$, $6 \cdot 10^8$ $K^\pm$ so far &
$\phi$ factory, pure $K$ &
2000+ \\
Protvino-U70 &
25 GeV/$c$, $> 10^9$ $K^-$ &
Unseparated ($\pi/K \approx 30$) & 
2001+ \\
CERN-SPS &
60 GeV/$c$, $< 10^{11}$ $K^\pm$ so far &
Unseparated, ($\pi/K \approx 10$) &
2003+ \\
\hline
Protvino-U70 &
12-18 GeV/$c$ & 
Separated ($\pi/K <2$) &
2004+ \\
FNAL-MI &
22 GeV/$c$ &
Separated ($\pi/K<0.5$) &
2007+ \\
J-PARC-PS &
600-700 MeV/$c$ &
Separated ($\pi/K <0.3$) &
2008+ \\
\hline
\end{tabular}
\label{tab:world}
\end{center}
\end{table*}

Among new facilities, one where a strong kaon physics program is foreseen is
the J-PARC 50 GeV proton synchrotron in construction at Tokai in Japan 
\cite{JPARC}.
This high-intensity machine ($2 \cdot 10^{14}$ p/3.4 s) is expected to 
deliver beam to experiments in 2008. Two beam lines are foreseen in the
experimental hall (only one at startup), one of which with a low energy 
($\sim$ 600 MeV/$c$) separated $K^+$ beam ($\sim 10^7$ $K^+$/s). 
Several letters of intent for charged kaon experiments have been submitted, 
including: an improved transverse muon polarization experiment to reach 
$\sim 1 \cdot 10^{-4}$ sensitivity, an extensive program for the complete 
measurement of $K^+$ decay modes, a dedicated $K^+_{e3}$ branching ratio 
measurement, studies of pionium and $\pi K$ atoms, and a $K^+ \rightarrow 
\pi^+ \nu \overline{\nu}$ measurement.
One can question which of these measurements will remain central in kaon 
physics by 2008, and the answer clearly depends also on the success of the 
ongoing experimental efforts. 

Considering the availability of charged kaons in the world, one is led to 
the picture depicted in table \ref{tab:world}; the sheer amount and quality 
of the ongoing and future activities, including the ones which were discussed 
at this workshop involving upgrades of existing machines and/or experiments, 
clearly indicate how the field is a very active one.

Playing the seer, one could guess the following possible scenario concerning 
physics with charged kaons at a time where the LHC is starting: all large 
branching ratios of the charged kaon will be known with high accuracy, with 
the universality tests which they imply, and including a consistent value of 
$|V_{us}|$ from the kaon sector; some CP-violating charge asymmetries will be 
measured at the $\sim 10^{-4}$ level (and several others at $10^{-3}$), chiral 
perturbation theory predictions for several rare decays will be put under 
stringent test, and the experimental results will be closer to the theoretical 
precision in $\pi\pi$ scattering lengths. 

\section{CONCLUSIONS}
After the impressive success of the first CP violation studies with $B$ mesons
in recent years, also this heavier system is now reaching the stage in which 
challenging precision measurements are needed, requiring high statistics and a 
detailed understanding of several hard theoretical issues.
It seems therefore worthy to look again with renewed interest at the relative 
advantages of experiments with kaons, which can offer an alternative and 
complementary view on some deep unanswered questions of physics.

It is a pleasure to thank the organizers of the workshop (and the conveners 
too) for the very pleasant and interesting time in the wonderful environment 
of Alghero.

\end{document}